\documentstyle[12pt]{article}
\begin{document}
\tolerance=5000
\def\pp{{\, \mid \hskip -1.5mm =}}
\def\cL{{\cal L}}
\def\be{\begin{equation}}
\def\ee{\end{equation}}
\def\bea{\begin{eqnarray}}
\def\eea{\end{eqnarray}}
\def\tr{{\rm tr}\, }
\def\nn{\nonumber \\}
\def\e{{\rm e}}
\def\D{{D \hskip -3mm /\,}}

\  \hfill 
\begin{minipage}{3.5cm}
January 2000 \\
\end{minipage}

\vfill

\begin{center}
{\Large\bf Quantum (in)stability of dilatonic AdS backgrounds and holographic
renormalization group with gravity}

\vfill

{\sc Shin'ichi Nojiri}$^\clubsuit$\footnote{nojiri@cc.nda.ac.jp}, 
{\sc Sergei D. Odintsov}$^\spadesuit$\footnote{
e-mail: odintsov@mail.tomsknet.ru}

{\sc  and Sergio Zerbini}$^\diamondsuit$\footnote{
e-mail: zerbini@science.unitn.it }

\vfill

\

{\sl $\clubsuit$ Department of Mathematics and Physics \\
National Defence Academy, 
Hashirimizu Yokosuka 239, JAPAN}

\ 

{\sl $\spadesuit$
Tomsk State Pedagogical University, 634041 Tomsk, RUSSIA}

\ 

{\sl $\diamondsuit$ Department of  Physics and Gruppo 
Collegato INFN\\
University of Trento 38050-Trento, ITALY}

\vfill

{\bf Abstract}

\end{center}

Stability of dilatonic AdS spaces due to quantum effects of 
dilaton coupled conformal matter is considered. When such 
spaces do not exist on classical level, their dynamical 
generation occurs. Explicit examples corresponding 
to quantum creation of d4 dilatonic AdS Universe and of 
d2 dilatonic AdS Black Hole (BH) are presented. 
Motivated by holographic RG, in the similar
approach the complete d5 effective action is discussed. 
The intermediate region where it is the sum of two parts: 
bulk (classical gravity) and boundary quantum action is 
investigated. The effective equations solution representing 
d5 AdS Universe with warp scale factor is found. 
Four-dimensional de Sitter or AdS world is generated on 
the boundary of such Universe as a result of quantum effects.



\newpage

\section{Introduction}

The interest in the study of questions related with AdS 
backgrounds is caused by several reasons.
 From one side, in AdS/CFT set-up \cite{AdS} the investigation 
of classical solutions of IIB SG with AdS sections may provide 
the information about boundary QFT in less dimensions. Second, 
AdS space has maximum number of Killing vectors and it is 
well-known candidate for vacuum state in various
SGs. For strings the AdS-like backgrounds are often suspected to 
be exact vacuum state. Third, some cosmological data indicate to 
the presence of spatial sections with negative curvature in the 
inflationary Universe. The recent numerous studies of warped
compactifications 
and brane cosmology have been iniciated by refs.\cite{RS}(the corresponding
literature is vast, see \cite{RS2} and refs. therein). That suggests the
possibility of AdS-like stage 
 which could be important for various reasons 
(hierarchy problem, etc) in the early Universe.

Reserving the possibility of dynamical generation of AdS 
regions at the early (multidimensional) Universe the question 
could arise: what are the mechanisms responsible for such process? 
In particular, if AdS BHs exist they should be presumbly produced 
as the primordial BHs at the very early Universe where
quantum effects play the dominant role. Hence, the natural 
mechanism for creation-annihilation of AdS backgrounds could 
be defined by quantum effects. Indeed,  it has been shown 
recently \cite{BO} that quantum anihilation of 4d AdS Universe 
typically occurs. However, the presence of dilaton 
may change the situation to quantum creation of dilatonic 
AdS Universe (subject to fine-tuning of theory parameters).

In the present work we investigate the role of quantum 
conformal matter effects (using large $N$, anomaly induced 
effective action) in the dynamical realization of
dilatonic AdS backgrounds in various dimensions. In the next section
the 4d dilatonic classical gravity with $N$ dilaton coupled 
quantum fermions is considered. From the solution of effective 
equations it follows that quantum corrected dilatonic AdS 
Universe may occur, even in situation where such classical 
solution was absent. However, the probability of generation 
of AdS Universe is significally less than of quantum creation 
of de Sitter Universe. It is interesting that some features 
of quantum creation of dilatonic AdS Universe should be also 
typical for the same process for primordial AdS BHs. To demonstrate 
this, in section three we present two-dimensional
exactly solvable model. It is the same dilatonic gravity with dilaton
coupled massless quantum fermions in two dimensions. The complete 
anomaly induced effective action is found. The quantum 2d 
dilatonic AdS BH which did not existed on classical level 
is constructed as the solution of the effective equations. 
In other words, the creation of quantum 2d AdS BH occurs.

 From another side, in AdS/CFT set-up where AdS background 
is introduced from the very beginning the question on 
realization of gravity in boundary world appears. Some progress 
in this direction is made within Randall-Sundrum scenario 
which may suggest the solution of hierarchy problem. In fourth 
section we discuss the holographic RG action leading to warped 
compactification (RS-like Universe) in the region where 
both sides of AdS/CFT correspondence (i.e. bulk and boundary) 
are still relevant. Then, using the same anomaly induced 
effective action of previous sections (boundary) we suggest 
the way it should appear in the dynamics of five-dimensional 
world on equal footing with 5d gravity action. As a result
the dynamical effective equations could be solved realizing 
5d AdS Universe with warp factor. On the boundary of such 
Universe the de Sitter (inflationary) world occurs. It is 
actually induced by quantum effects (not four-dimensional 
cosmological constant introduced on brane by hands).
In the final section we present short resume and mention 
some open problems.

\section{Quantum instability of 4d AdS dilaton universe}

 Let us consider the  4-dimensional dilaton gravity theory with the 
following action:
\be
S=\int d^4x \sqrt{-g} \left[ 
 -\frac{1}{16 \pi G}(R+ 6\Lambda) +\alpha
(\nabla_\mu \phi)(\nabla^\mu \phi) \right]\,,
\label{dg4}
\ee
where $\Lambda$ is the cosmological constant and $\alpha$ some suitable 
parameter. For constant dilaton $\phi$ and negative cosmological constant 
the classical background solution corresponds to the 4d AdS space. 
Even for non-constant dilaton, there are solutions interpolating 
between asymptotically AdS and flat space with singular dilaton \cite{NO1}. 
Our primary 
interest will be in (asymptotically) AdS background for the theory 
defined by (\ref{dg4}).

Few remarks are in order. There are different interpretations of action 
(\ref{dg4}). First of all, this action may be considered as the 
compactification of the II B supergravity bosonic sector (say, a Freund-Rubin
ansatz and background like $S_5 \times S_1 \times AdS_4$).

Second, making a conformal transformation   of the metric and a suitable 
redefinition of the dilaton, one arrives at the Brans-Dicke theory  
(for a review, see \cite{will}), with non-trivial dilatonic potential. 
Hence, the action (\ref{dg4}) may be considered as a Brans-Dicke theory in 
the Einstein frame, which has been  argued to be the physical one 
(for a review, see \cite{FGT}).

We will be interested in the investigation of the stability of the
(classical) 
AdS background in the theory (\ref{dg4}), under the quantum fluctuations of
the 
conformal matter. As  matter Lagrangian, we take the one which corresponds
to $N$ massless dilaton coupled Dirac spinors
\be
\label{dirac}
L_M=\exp{(A\phi)}\sum_{i=1}^N \bar\psi_i\gamma^\mu\nabla_\mu\psi^i\ .
\ee
Here $A$ is some constant parameter.
The above strong matter-dilaton coupling is typical  for any matter- 
Brans-Dicke theory  in the Einstein frame. Transforming  the variables 
in the theory  (\ref{dg4}) to the Brans-Dicke gravity in the Jordan frame, one
can easily show that $L_M$  is tranformed into the usual spinor Lagrangian
with 
no dilaton coupling. That is why $L_M$ is just how the matter looks in the 
Brans-Dicke gravity within the Einstein frame. Additional motivations to 
consider the dilaton coupled matter Lagrangian as in  (\ref{dirac}) come 
from the supergravity side, where such coupling is a common feature when the 
dilaton is present in the supergravity multiplet. 
The number of spinors $N$ is to be considered very large, in order to justify
why one can neglect the proper quantum gravity contributions. 

The quantum effective action for dilaton coupled spinor has been found in 
ref. \cite{NNO} by integrating the conformal anomaly. This quantum effective 
action should be added to the classical one $S$ (there is only the 
dilaton-gravitational background under consideration).

Let us now define the space-time we are going to work with. 
We consider the 4d AdS background with the static metric 
\be
ds^2=\e^{-2\lambda \tilde{x}_3} \Big[dt^2-(dx_1)^2-(dx_2)^2 \Big]-
(d\tilde{x}_3)^2\,.
\label{3}
\ee
It has a  negative cosmological constant $\Lambda=-\lambda^2$. Making the 
coordinates transformations 
\be
y=\frac{\e^{\lambda \tilde{x}_3}}{\lambda} 
\ee
one can present (\ref{3}) in the conformally flat form  
\be
ds^2=a(y)^2 \eta_{\mu \nu}dx^{\mu}dx^{\nu}\,,
\label{33}
\ee
where
\be
a=\e^{-\lambda \tilde{x}_3}=\frac{1}{\lambda y}\,.
\ee
This form of the metric  is very useful in the AdS/CFT correspondence. It is 
also clear that the classical AdS Universe may be trivially realized as 
solution of the theory (\ref{dg4}), with negative cosmological constant and
constant dilaton.

Here our purpose will be the investigation of the role of the quantum effects 
to the dilaton AdS universe. To this aim, we shall consider the 
metric  (\ref{33})  with an arbitrary scale factor to be determined 
dynamically.
The anomaly induced effective action of ref. \cite{NNO} on such background 
may be written in the following form \cite{GOZ}:

\be
\label{1ef}
W=V_3\int dy \left\{2b_1 \sigma_1 \sigma_1''''
 - 2(b + b_1)\left( \sigma_1'' - {\sigma_1'}^2 \right)^2\right\}\ .
\label{7}
\ee
Here, $V_3$ is the (infinite) volume of 3-dimensional flat space-time( time
is included), 
 $\sigma=\ln a(y)$, $\sigma_1=\sigma+\frac{A\phi}{3}$,
$'\equiv{d / dy}$, and
$b={3N \over 60(4\pi)^2}$, $b_1=-{11 N \over 360 (4\pi)^2}$.
It should be noted that (\ref{7}) may be regarded as the complete one-loop 
effective action.
The classical  gravitational action on the background (\ref{3}) with 
non-trivial dilaton may be 
obtained from (\ref{dg4}) and reads
\be
S=V_3 \int dy \left\{ \frac{1}{\chi}\left[ 6(\sigma''+{\sigma'}^2) 
\e^{2\sigma}-6\Lambda \e^{4\sigma} \right]+          
\alpha {\phi'}^2 \e^{2\sigma} \right\}\,.
\label{ca}
\ee
with $\chi=16 \pi G$. The quantum corrections can be taken into account 
starting from the effective equations obtained from the effective action
$S+W$. These equations look similarly to the ones  associated with the 
De Sitter quantum corrected universe \cite{GOZ}, with the addition of the 
cosmological constant contribution and an opposite sign in the Einstein term:

\bea
&&\tilde{C} \e^{({A \phi}/{3})}-\frac{12}{\chi}a''-\frac{24\Lambda a^3}{\chi}
 +2\alpha a\phi'^2=0 \nn 
&& \frac{A}{3}\tilde{C} a \e^{({A \phi}/{3})}-2\alpha (a^2 \phi')'=0\,,
\label{8}
\eea
where
\be
\tilde{C}=
 -\frac{4b}{\tilde{a}} \left[ \frac{\tilde{a}''''}{\tilde{a}}
 -\frac{4\tilde{a}' \tilde{a}'''}{\tilde{a}^2} 
 - \frac{3\tilde{a}''^2}{\tilde{a}^2}\right] 
 -\frac{24}{\tilde{a}^4} \left[ (b-b_1) \tilde{a}'' \tilde{a}'^2 
 + b_1\frac{\tilde{a}'^4}{\tilde{a}} \right] \,, 
\label{88}
\ee
and
\be
\tilde{a}=a  \e^{({A \phi}/{3})}\,. 
\ee
First, we note that in absence of dilaton and quantum corrections 
$(\alpha=b=b_1=0)$, there exists a special solution 
\be
a(y)=\frac{1}{\sqrt{-\Lambda}y}\,,
\ee
corresponding to the AdS Universe, namely such universe exists at classical 
level.

Furthermore, in the absence of the dilaton and vanishing cosmological
constant,
only the first of (\ref{8}) survives. It may be solved via the special 
ansatz $a=c/y$ with the constraint $c^2=b_1\chi$. Since $b_1 <0$, one gets
an imaginary scale factor $a$, namely a quantum annihilation of the AdS 
universe, as it was shown in detail  in ref. \cite{BO}. Hence, classical AdS 
Universe is not stable under the action of quantum fluctuations. Note that 
dilaton is constant there.

Now, one can ask what happens  in the general situation within 
dilatonic gravity with non-constant dilaton with regard to the AdS universe. 
In other words, there is the question about the stability of the AdS universe 
in II B Supergravity under the action of the quantum matter fluctuations.  

The general solution of the system of differential equations (\ref{8}) is 
very difficult to find. Apart from numerical studies, the best one can do
is to 
search for special solutions. Motivated by the similar dilatonic De Sitter 
universe solution \cite{GOZ}, one may try 
\be
a(y)=\frac{1}{H y}\,,\,\,\,\,\,\, 
\phi'(y)=\frac{1}{H_1 y}\,,
\label{10}
\ee
where $H$ and $H_1$ are some constants to be determined.
 From Eqs. (\ref{8}) one obtains
\be
\frac{aA}{3}\left[ \frac{12a''}{\chi}+\frac{24\Lambda a^3}{\chi}
 -2\alpha a \phi'^2\right]-2\alpha (a^2 \phi')'=0\,.
\label{12}
\ee

Substituting Eqs.~(\ref{10}) in the second of (\ref{8}) and 
(\ref{12}), we obtain
\be
\frac{12 \Lambda}{\chi H^2}= \frac{ \alpha}{H_1^2}
 -\frac{9 \alpha}{A H_1}-\frac{12}{\chi}\,. 
\label{11}
\ee
\be
\frac{81 \alpha}{2A b H^2}= -(A-3H_1)
\left(\frac{18H_1^2- 2\tilde{b}(A-3H_1)^2 }{H_1^2}\right)\,, 
\label{16}
\ee
with $\tilde{b}=1+\frac{b_1}{b}$. As a result, one may eliminate $H$ from the 
first equation and obtain a third  order complete algebraic equation for 
the quantity $H_1$, which in principle can be solved.
However, for $\Lambda=0$, there is the complete decoupling of the two 
equations and one easily arrives  at
\be
H_1=\frac{1}{24}\left[ -\frac{9 \alpha \chi}{A}\pm 
\sqrt{\frac{81 \alpha^2 \chi^2}{A^2}+ 48 \alpha \chi}\right]\,,
\label{17}
\ee 
and the other quantity $H$ may be obtained from equation (\ref{16}): 
\be
\label{18}
{1 \over H^2}={2b \over 81\alpha}
\left[\left(-18 + 81 \tilde b\right)A^2 
+ {24 \tilde b A^4 \over \alpha \chi} \pm {\alpha \chi \over 3\tilde b A}
\sqrt{\frac{81 \alpha^2 \chi^2}{A^2}+ 48 \alpha \chi}\right]\ .
\ee
Since $\tilde b={7 \over 11}>0$ and $-18 + 81 \tilde b>0$, there is 
always a real solution for $H$ at least for positive $\alpha$ if we 
choose the sign $\pm$ in front of the square root properly. 

Hence, we demonstrated that in presence of non-constant dilaton 
the quantum AdS Universe solution in dilatonic gravity is less unstable.
At least, it may be realized while it didnt exist on classical level!
However, it is less stable than corresponding de Sitter Universe
\cite{GOZ} as it may be easily seen from the explicit solution.
In fact, the mechanism presented in this section may serve as
the one corresponding to quantum creation of primordial AdS Black 
Holes in early Universe. However,  for 4d AdS BH one should calculate
the extra piece of effective action which is non-local and very complicated.
The complete calculation of it is not known, presumbly it could be found 
only as expansion on theory parameters. That is the reason we prefer 
to present such analysis only in two dimensions.  

\section{ Quantum creation of 2d AdS black hole}

In this section we investigate the possibility for quantum creation
of 2d AdS BHs using methods developed in previous section.
Motivated by the 4-dimensional case, we may assume that the classical 
action  for the 2-dimensional dilaton gravity theory reads
\be
S=\int d^2x \sqrt{-g(x)} \left[ -\frac{R+6\Lambda}{\chi}+\frac{1}{2}
(\nabla_\mu \phi)(\nabla^\mu \phi)+
\exp{(A \phi)L_M} \right]\,,
\label{s2}
\ee
where $A$ is a constant parameter  and the matter 
Lagrangian is the one of two-dimensional Majorana spinors:
\be
\label{i}
L_M=\sum_{i=1}^N \bar\psi_i\gamma^\mu\nabla_\mu\psi^i\ .
\ee
Let us neglect the classical matter contribution since we are interested 
only in the one-loop EA induced by the conformal anomaly of the quantum 
matter. 

In two dimensions, a general static   metric  may be written in  the form
\be
ds^2=V(r)dt^2-\frac{1}{W(r)} dr^2\,.
\label{st}
\ee
It is well know that introducing the new radial coordinate $r^*$, defined by
\be
r^*=\int \frac{dr}{\sqrt{V(r)W(r)}}\,,
\ee
one gets a conformally flat space-time, 
\be
g_{\mu \nu}=V(r(r^*))\eta_{\mu \nu}=\e^{2\sigma(r^*)}\eta_{\mu \nu}\,.
\ee
We also assume that the field  $\phi$ depends only on 
$r^*$.

Since the conformal anomaly for the dilaton coupled 
spinor is  (see \cite{NNO})
\be
T={c \over 2}\tilde R \ ,
\ee
where $c={N /( 12\pi)}$ and $\tilde R$ is calculated on the 
metric $\tilde g_{\mu\nu}=\e^{2A\phi}g_{\mu\nu}$, 
the anomaly induced EA  in the local, non-covariant form  reads 
\be
\label{eas}
W={c \over 2}\int d^2x \tilde\sigma \tilde\sigma'' \ .
\ee
Here $\tilde\sigma = \sigma + A\phi$ and $'=\frac{d}{dr^*}$. 
Note that this is, up to a
non--essential constant, an exact one-loop expression.  
The total one-loop effective action is $S+W$, i.e.
\be
\label{tot}
S+W=
V_1\int dr^*  \left\{\frac{k_G}{2} 
\left(\sigma''-6\Lambda\e^{2\sigma}\right)+\frac{1}{2} (\phi')^2 
+\frac{c}{2}\left(\sigma + A\phi\right) 
\left(\sigma'' + A\phi''\right)  \right\}\ \,,
\ee
where $k_G=\frac{1}{8 \pi G}$, and  $V_1$ the (infinite) 
temporal volume. 
Since 2d Einstein theory is trivial, the whole dynamics 
appears as a result of quantum effects.
The equations of motion given by the variations of $\phi$
and $\sigma$ are 
\bea
\label{eqm1}
0&=&-\left(1 - cA^2\right)\phi'' + cA \sigma'' \\
\label{eqm2}
0&=& c\left( \sigma'' + A \phi''\right) - 6k_G \Lambda 
\e^{2\sigma}\ .
\eea
 From (\ref{eqm1}) and (\ref{eqm2}), we obtain
\be
\label{eqm3}
0= - 6k_G\Lambda \e^{2\sigma} 
+ {c \over 1 - cA^2 }\sigma''\ ,
\ee
which gives a constant $E$ of motion
\be
\label{eqm4}
E=-3k_G\Lambda \e^{2\sigma} 
+ {c \over 2\left(1 - cA^2 \right)}\left(\sigma'\right)^2\ .
\ee
We now change the coordinate by 
\be
\label{eqm5}
r=\int \e^{2\sigma} dr^*\ ,
\ee
which gives 
\be
\label{eqm6}
W(r)=V(r)=\e^{2\sigma}\ .
\ee
Then Equation (\ref{eqm4}) can be integrated to give
\bea
\label{eqm7}
&&\e^{2\sigma}={(r-r_0)^2 \over b} - a \nn
&& b\equiv {6(1-cA^2)k_G \Lambda \over c}\ ,
\quad a\equiv {E \over 3 k_G \Lambda}\ .
\eea
Here $r_0$ is a constant of the integration. 
If we further redefine, 
\be
\label{eqm8}
{1 \over l^2} \equiv b \ ,\quad cM \equiv {2r_0 \over b}
\ ,\quad k \equiv -a + {r_0^2 \over b}\ ,
\ee
we obtain a generic 2d AdS black hole solution, 
\be
W(r)=V(r)=\e^{2\sigma}=k-cMr+\frac{r^2}{l^2}
\label{ads2}
\ee
where $M$ may be interpreted as the mass of the BH. We may take 
$k=\pm 1$, or $k=0$. In general, we have a simple positive root, 
interpreted as horizon radius. In the case $k=1$ one must 
have $cM >2$. It is easy to show that the above metric has 
a negative constant scalar curvature and for large $r$, 
$V(r) \simeq \frac{r^2}{l^2}$, namely one gets the AdS 
asymptotic behavior.  For the sake of simplicity, let us 
consider the case $k=0$. 
In this case the horizon radius and the Hawking temperature read
\be
\label{rHbH}
r_H=cMl^2\,\,\,, \beta_H=\frac{4 \pi}{cM}=\frac{4 \pi l^2}{r_H}\,. 
\ee
Since
\be
r^*=\frac{l^2}{r_H}\ln{ \frac{r-r_H}{r}}\,,
\ee
the old radial coordinate as a function of the new one is
\be
r=\frac{r_H}{1-\e^{4\pi T_H r^*  }}\,.
\ee
Note that the horizon $r=r_H$ corresponds to $r^*$ going 
to $- \infty$. The $\sigma$ function is given by
\be
\sigma(r^*)=\ln \frac{r_H}{l}-\ln(1-\e^{4\pi T_H r^*})+2\pi T_H r^*\,.    
\ee
 From (\ref{eqm1}), one obtains the following first integral of motion
\be
 -\left(1 - cA^2\right) \phi'(r^*)+c A\sigma'(r^*)=c_1\,,
\ee
where $c_1$ is an integration constant. Thus the
solution for the dilaton field is
\be
\phi(r^*)=\phi_0+c_1r^*
+ {c A \over 1 - cA^2}\sigma(r^*)\,,
\ee
where $c_1$ is an   integration constant.
As a function of the old radial coordinate one has
\be
\phi(r)=\phi_0+c_1 \ln \frac{r-r_H}{r}
+ {c A \over 1 - cA^2}\ln \frac{\sqrt{r(r-r_H)}}{l} \,.
\ee
Note that the quantum correction to the dilaton diverges 
on the horizon.

Thus, using anomaly induced effective action for dilaton coupled 
spinor we proved the possibility of quantum realization of 2d 
AdS BH which did not exist on classical level. It is interesting 
that unlike to 4d case where EA for AdS BH is not completely known, 
2d case is exactly solvable. Our solution may be interpreted as 
quantum creation of dilatonic AdS BH.

The finite Hawking temperature in (\ref{rHbH}) tells that the 
Hawking radiation (for a recent review, see \cite{KV}) should occur for our
BH solution. If we consider 
the time dependent perturbation, one could trace the fate of the 
black hole. This will be discussed elsewhere.

\section{Holographic Renormalization Group and Dynamical Gravity}

In the standard AdS/CFT correspondence \cite{AdS} one can think 
about the simultaneous incorporation of string compactification 
with exponential warp factor (Randall-Sundrum compactification 
\cite{RS,RS2}) and the holographic map between
5d supergravity and 4d boundary (gauge) theory. Moreover, 
it could be extremely interesting to do it in such a way 
that dynamical gravity would appear on the boundary side. 
One possibility to realize such a mechanism is
presumbly related with holographic renormalization group (RG), see
\cite{BK,VV} for an introduction. Probably the most important 
element of such RG (as well as in holographic correspondence 
between 5d SG and 4d dual gauge theory) is the identification 
of fifth AdS coordinate with the RG parameter of 4d boundary 
theory. In particular, this gives the way to study RG flows in 
4d gauge theory via the investigation of classical AdS-like 
solutions of 5d gauged SGs.

There was very interesting suggestion in this respect 
in ref.\cite{VV} to consider low-energy effective 
action (EA) in the region where field theoretical 
quantities and analogous supergravity quantities 
could be considered on equal foot. In other 
words, this is the way to match two dual 
descriptions into the global picture of some, 
more universal RG flow. Immediate consequence of 
such point of view is the possible explanation of 
smallness of cosmological constant, the stability of 
flat spacetime along the RG flow and possible 
understanding of 4d gravity appearence in standard 
AdS/CFT set-up.

Our purpose will be to find the explicit 
realization of ideas of ref.\cite{VV} via the construction 
of the corresponding phenomenological model. First of all, it will be 
necessary to repeat two starting points of the  
consideration in ref.\cite{VV}. 

In the calculation of complete low-energy EA in  
AdS/CFT set-up one can divide it into a 
high energy and low energy pieces, separated by 
some given RG scale (fixed value of radial 
coordinate):
\be
\label{I1}
S=S_{\rm UV} + S_{\rm IR}\ .
\ee
Here $S_{\rm UV}$ is obtained from the original stringy 
action as a result of specific compactification. 
$S_{\rm IR}$ may be identified with the quantum effective 
action of (gauge and matter) low energy theory. 

Let us start now the explicit construction 
of the model. Consider 5d warped AdS metric: 
\be
\label{I2}
ds^2 = a_1^2(r)\tilde g_{\alpha\beta}dx^\alpha dx^\beta 
 - dr^2
\ee
where $r$ is radius of d5 AdS, or RS Universe,
i.e., $a_1 =a_1(r)$ is scale factor of d5 AdS and 
$a_1(r)$ usually depends exponentially on the 
radial coordinate. 
$\tilde g_{\alpha\beta}$ is 4d metric of boundary, 
time dependent FRW Universe. We assume that 
$\tilde g_{\alpha\beta}=a^2(\eta)\eta_{\alpha\beta}$ 
where $\eta$ is conformal time and 
$\eta_{\alpha\beta}$ is 4d Minkowski tensor. 
As $\tilde g_{\mu\nu}$ corresponds to conformally flat space,
it is defined by conformal time dependent scale factor $a$. 
Hence we have two scale factors.
 One can discuss now 
the structure of low-energy effective action. 
Truncation of $S_{\rm UV}$ gives basically the bosonic sector 
of 5d gauged supergravity (for simplicity, 
we consider the situation with only one scalar (dilaton) 
which is quite typical): 
\be
\label{I3}
S_{\rm UV}=\int d^5x \sqrt{-g_{(5)}}\left\{V(\phi) 
- {1 \over H(\phi)} R_{(5)} + V_1(\phi) \nabla_\mu\phi 
\nabla^\mu\phi\right\}\ .
\ee
At first step, we limit ourselves to even simpler 
situation of constant dilaton. (Note that holographic 
RG implies that $\phi$ corresponds to the coupling constant of dual QFT).
 The reason 
is that we are searching for 4d dynamic gravity, at 
least qualitatively. Then,
\be
\label{I4}
S_{\rm UV}=\int d^5x\sqrt{-g_{(5)}}\left\{-{1 \over H}R_{(5)} 
 - {6\Lambda \over H}\right\}
\ee
where $V(\phi=\mbox{const})\equiv {6\Lambda \over H}$. 
We consider 5d AdS background with some 4d time-dependent 
conformally-flat boundary in this theory 
as vacuum state. The question 
is: can boundary quantum effects (instead of 4d cosmological constant)
stabilize such space?

The 4d quantum effective action of low-energy theory on 
the conformaly-flat space $g_{\mu\nu}=a^2(\eta)\eta_{\mu\nu}$ 
looks as (see, for example, \cite{BO})
\be
\label{I5}
W=V_3\int d\eta \left\{2b_1 \sigma \sigma''''
 - 2 (b + b_1)\left(\sigma'' - {\sigma'}^2\right)^2\right\}
\ee
where $\sigma=\ln a(\eta)$, $V_3$ is space volume, 
$\sigma'={d \sigma \over d\eta}$, for ${\cal N}=4$ 
$SU(N)$ SYM theory $b={N^2 -1 \over 4(4\pi )^2}$, $b_1=-b$. 
In general, $b>0$, $b_1<0$ and $b\neq b_1$. 

One has to relate $W$ with $S_{IR}$. We 
consider d5 AdS background with the metric of the form:
\be
\label{I6}
ds_5^2 = -dr^2 + a_1^2(r)a^2(\eta)\eta_{\mu\nu}dx^\mu 
dx^\nu\ .
\ee

One knows that $S_{\rm IR}=W$ at $r=r_0$, cut-off 
scale. On the other side in AdS limit the 
description is completely from supergravity 
side. So, at $r\rightarrow r_A$, $S_{\rm IR}\rightarrow 0$. 
Then one can adopt the phenomenological 
approach where 
\be
\label{I7}
S_{\rm IR}=\int f\left(a_1(r)\right) W dr
\ee
so that $f\left(a_1(r)\right)$ satisfies above relations 
connecting $S_{\rm IR}$ and $W$ . Some choice for it
 is explicitly given below. 
Then, one can solve Eqs. of motion from 
$S_{\rm UV} + S_{\rm IR}$ on the background (\ref{I6}). 

Under the assumption of the metric in (\ref{I6}), the sum of 
the actions (\ref{I4}) and (\ref{I7}) has the following 
form 
\bea
\label{SS1}
&& S_{\rm UV} + S_{\rm IR} \nn
&& = V_3 \int dr\, d\eta\Bigl[
 -\Bigl\{\e^{2\varphi + 2 \sigma}\left(6\sigma_{,\eta\eta} 
+ 6\sigma_{,\eta}^2\right) \nn
&& \left. \ + \e^{4\varphi + 4\sigma}\left( -8 \varphi_{,rr} 
 - 20 \varphi_{,r}^2 - {16 \over l}\varphi_{,r} \right)\right\}
{1 \over H} \nn
&& \  - {6\Lambda \over H}\e^{4\varphi + 4\sigma} 
+ f(\varphi)\left\{2b_1\sigma_{,\eta\eta}^2 
 - 2(b + b_1)\left(\sigma_{,\eta\eta} - \sigma_{,\eta}^2
\right)^2\right\} \nn
&& \left.\ + {A \over 4lH}\left(\e^{4\varphi + 4\sigma}\right)_{,r} 
+ {B \over H}\left(\e^{2\varphi + 2 \sigma}\sigma_{,\eta}
\right)_{,\eta}\right]\ .
\eea
Here $\varphi=\ln a_1(r)$, 
${2 \over l^2}=\Lambda$ and $\cdot_{,r}={d \cdot \over dr}$, 
$\cdot_{,\eta}={d \cdot \over d\eta}$. We add the total 
derivative terms proportional to constant parameters $A$ and $B$ 
and we also rewrite the term proportional to $b_1$ by using 
the total derivative. We should note, however, that the total 
derivative terms are not irrelevant to the equations of motion 
as we show it explicitly later, which might be caused by the fact 
that the 5d general covariance is not always kept in (\ref{I7}). 
The parameters $A$ and $B$ can be later determined by the 
consistency conditions. 
The equations of motion given by the variation over $\sigma$ and 
$\varphi$ have the following forms:
\bea
\label{SS2}
0&=&-{1 \over H}\e^{2\varphi + 2 \sigma}
\left(12\sigma_{,\eta\eta} + 12\sigma_{,\eta}^2\right) \nn
&& + {4 \over H}\e^{4\varphi + 4\sigma}\left( -8 \varphi_{,rr} 
 - 20 \varphi_{,r}^2 - {16 \over l}\varphi_{,r} 
+ {A \over l}\varphi_{,r} - 6\Lambda\right) \nn
&& + 4f(\varphi)\left\{4b_1\sigma_{,\eta\eta\eta\eta} 
 - 2(b + b_1)\left(2\sigma_{,\eta\eta\eta\eta} 
 - 12\sigma_{,\eta}^2 \sigma_{,\eta\eta}  \right)\right\} \\
\label{SS3}
0&=&-{1 \over H}\e^{2\varphi + 2 \sigma}
\left\{6\sigma_{,\eta\eta} + 6\sigma_{,\eta}^2
 -B\left(\sigma_{,\eta\eta} + 2 \sigma_{,\eta}^2\right) \right\} \nn
&& + {1 \over H}\e^{4\varphi + 4\sigma}\left( -24 \varphi_{,rr} 
 - 48 \varphi_{,r}^2 - 24\Lambda\right) \nn
&& + f'(\varphi)\left\{2b_1\sigma_{,\eta\eta}^2 
 - 2(b + b_1)\left(\sigma_{,\eta\eta} - \sigma_{,\eta}^2
\right)^2\right\} \ .
\eea
Multiplying $\e^{-4\sigma}$ and differentiating it with 
respect to $\eta$, one gets
\bea
\label{SS4}
0&=&-{1 \over H}\e^{2 \sigma}{d \over d\eta}\left\{\e^{-2\sigma}
\left(12\sigma_{,\eta\eta} + 12\sigma_{,\eta}^2\right)\right\} \nn
&& + 4f(\varphi){d \over d\eta}\left[
\left\{4b_1\sigma_{,\eta\eta\eta\eta} 
 - 2(b + b_1)\left(2\sigma_{,\eta\eta\eta\eta} 
 - 12\sigma_{,\eta}^2 \sigma_{,\eta\eta}  \right)\right\}\right]\ .
\eea
In order that Eq.(\ref{SS4}) has some reasonable solution,
we choose that 
$f(\varphi)$ should be proportional 
to $\e^{2\varphi}$:
\be
\label{SS5}
f(\varphi)=\e^{2\left(\varphi - \varphi_0\right)}\ .
\ee
Then UV limit, where $f=0$, corresponds to $\varphi=-\infty$ and 
IR limit, where $f=1$, corresponds to $\varphi=\varphi_0$. 
The value of $\varphi$ is determined later from the equations 
of motion. 
Assuming 
\be
\label{SS6}
\sigma=-\ln \eta \ ,\quad \varphi={a \over l}r\ ,\quad 
\mbox{($a$ is a constant)}
\ee
one obtains the following from the equations of 
motion (\ref{SS2}) and (\ref{SS3}) :
\bea
\label{SS7}
0&=& \left[-{24 \over H} + 4\cdot24\e^{-2\varphi_0}
 b_1 \right]{1 \over \eta^4} \nn
&& + {4 \over l^2 H}\e^{{2ar \over l}}\left( -20 a^2 
 -16a + Aa + 72 \right) \\
\label{SS8}
0&=& \left[{-24 + 6B \over H} + 4\e^{-2\varphi_0} b_1
\right]{1 \over \eta^4} \nn
&& + {48 \over l^2 H}\e^{{2ar \over l}}\left( - a^2 + 1 \right)\ .
\eea
Then the solution is 
\bea
\label{SS9}
&& \e^{2\varphi_0}= 4 H b_1 \ ,\quad 
B= {23 \over 6} \nn
&& (a, A)=(1,-32)\ \mbox{or}\ (-1, 68)\ .
\eea
Then the metric  is given by
\be
\label{SS10}
ds^2 = -dr^2 + {\e^{\pm {2(r-r_0) \over l}} \over \eta^2}
\left(d\eta^2 - \sum_{i=1}^3 \left(dx^i\right)^2\right)\ .
\ee
Here $\pm {r_0 \over l}=\varphi_0$.
The metric of the wall of the brane is  
\be
\label{dS}
ds_{\rm wall}^2={1 \over \eta^2}
\left(d\eta^2 - \sum_{i=1}^3 \left(dx^i\right)^2\right)\ .
\ee
If one changes the time variable $\eta$ by $\eta=\e^{-t}$, we obtain 
\be
\label{dS2}
ds_{\rm wall}^2=dt^2 - \e^{2t}
\sum_{i=1}^3 \left(dx^i\right)^2\ ,
\ee
It is nothing 
but that of de Sitter space, 
which can be regarded as inflationary universe. It is interesting that 
Hubble parameter is depending from radial coordinate of 5d AdS Universe.
Therefore we have obtained the time dependent solution in the form of  
warped compactification, which is caused by the quantum correction 
coming from the boundary QFT. 
In the above treatment, we have assumed that the wall lies at 
$r=r_0$ since $f=1$ there. We need, however, to check 
the dynamics of the wall by solving junction equation coming 
from the surface counterterm, which should include $W$ 
in (\ref{I5}) as a quantum correction. 

If we assume that the metric has the following form instead of 
(\ref{I6}), 
\be
\label{I6b}
ds_5^2 = -dr^2 + a_1^2(r)a^2(y)\left( dt^2 - dy^2 - 
\left(dx^1\right)^2 - \left(dx^2\right)^2 \right)  ,
\ee
one finds $\e^{2\varphi_0}$ in (\ref{SS5}) is 
\be
\label{SS9b}
\e^{2\varphi_0}= - 4H b_1 \ ,
\ee
instead of (\ref{SS9}). Since $b_1>0$ in most of cases, 
Eq.(\ref{SS9b}) seems to be inconsistent. 
In case of $b_1 <0$, however, one obtains the 
following metric
\be
\label{SS10b}
ds^2 = - dr^2 + {\e^{\pm {2(r-r_0) \over l}} \over y^2}
\left(dt^2 - dy^2 - \left(dx^1\right)^2 
- \left(dx^2\right)^2 \right)\ .
\ee
Then the metric of the wall of the brane is given by 
\be
\label{AdS}
ds_{\rm wall}^2={1 \over y^2}
\left( dt^2 - dy^2 - \left(dx^1\right)^2 
- \left(dx^2\right)^2 \right)\ .
\ee
The metric in (\ref{AdS}) is nothing but that of 4d AdS. Hence, one can get 
5d AdS Universe with warp scale factor a la Randall-Sundrum where 
4d AdS world is generated on the wall. Again, as in section 2 the
probability of realization of 4d AdS is less than the one for de Sitter
Universe.  

Hence, we presented the model where warped RS type scenario may 
be realized simultaneously with generation of inflationary 
Universe (or less stable AdS) on the wall. Dynamical 4d gravity is induced
from
background gravitational field on the boundary.
The source for such mechanism is quantum effects due to boundary 
QFT. It is not quite clear how one can estimate exactly these 
quantum effects.
That is the reason we adopted the phenomenological approach 
introducing some cut-off, interpolating, fifth coordinate 
dependent  function in such a way that near AdS the theory is 
described by 5d SG.  Far away of AdS,
at some fixed radius it is described by anomaly induced effective 
action of dual 4d QFT. There is, of course, some ambiguity in 
the choice of this function. However, that may be considered as 
kind of usual regularization dependence 
in frames of holographic RG.

\section{Discussion}

In the present work the dynamical generation of AdS backgrounds 
in dilatonic gravity with quantum dilaton coupled matter is discussed.
The dynamical generation is caused by quantum effects which we 
incorporate via using the anomaly induced effective action. 
It is shown that in such a way the 2d dilatonic AdS BH as well 
as 4d AdS Universe may be created. Hence, via effective action 
approach the account of quantum corrections gives rise the possibility to 
create the primordial AdS BHs or primordial  regions 
with negative curvature in the early Universe. 
Of course,
the probability to induce such spaces is normally less than the
corresponding one for de Sitter regions or de Sitter-like BHs. 

Holographic RG is also considered. The holographic effective action 
is taken in the intermediate region where it consists of two parts: 
UV (bulk, i.e. 5d classical gravity) and
IR (boundary QFT contribution derived via anomaly induced 
effective action). These two terms are of the same order. 
The solution of effective equations 
suggests that one can realize dynamically 
(i.e. thanks to quantum effects) 5d AdS Universe with warp 
scale factor where 4d boundary is inflationary Universe. 
Less stable boundary 4d AdS world could also occur.

There are different ways to generalize the results of this work. 
First of all, one can study in detail the properties of 
dynamically generated AdS backgrounds, say, Hawking radiation 
in AdS BH, realization of 4d AdS BH, etc. More general backgrounds 
which are asymptotically AdS ones may be found also but only 
numerically. Second, it would be interesting to generalize the 
results of section four to the case of non-constant dilaton. 
However in that case the role of interpolating, cut-off function 
should be better understood. Or, another way to introduce the 5d 
IR effective action due to boundary QFT should
be presented. In any case, the introduction of background gravity 
via anomaly induced effective action on the boundary leads to 
appearence of dynamical gravity in self-consistent way, via solution of
effective 
equations. Clearly, that such mechanism may be realized also 
for another dimensions. These questions will be discussed elsewhere.

\ 

\noindent
{\bf Acknowledgments}.
The research by SDO has been supported in part by RFBR Grant
N99-02-16617.
The research by SZ and SDO has been supported in part by INFN,
Gruppo Collegato di Trento.

\end{document}